\documentclass[a4paper,twocolumn]{emulateapj}
\pdfoutput=1
\usepackage{amstext}
\usepackage{apjfonts}
\usepackage{amsmath}
\usepackage{amssymb}
\usepackage{natbib}

\usepackage[colorlinks=true,linkcolor=blue,citecolor=blue]{hyperref}

\begin{document}

\title{Compact Stellar Binary Assembly in the First Nuclear Star Clusters  and \lowercase{$r$}-Process Synthesis in the Early Universe}

\author{Enrico Ramirez-Ruiz\altaffilmark{1,2,3}, Michele
  Trenti\altaffilmark{4,5}, Morgan MacLeod\altaffilmark{1}, Luke
  F. Roberts\altaffilmark{6,8}, William H. Lee\altaffilmark{7} and Martha
  I. Saladino-Rosas\altaffilmark{7}}

\altaffiltext{1}{Department of Astronomy and Astrophysics, University of California, Santa Cruz, CA 95064}
\altaffiltext{2}{Radcliffe Institute for Advanced Study, Harvard University, Cambridge, MA 02138}
\altaffiltext{3}{Niels Bohr Institute, University of Copenhagen, DK-2100 Copenhagen, Denmark}
\altaffiltext{4}{Kavli Institute for Cosmology and Institute of Astronomy, University of Cambridge, Madingley Road, Cambridge CB3 0HA, UK}
\altaffiltext{5}{School of Physics, University of Melbourne, VIC 3010, Australia}
\altaffiltext{6}{TAPIR, California Institute of Technology, Pasadena, California 91125, USA}
\altaffiltext{7}{Instituto de Astronom\'ia, Universidad Nacional Aut\'onoma de M\'exico, M\'exico DF 04510, M\'exico}
\altaffiltext{8}{NASA Einstein Fellow}

\begin{abstract} 
Investigations  of elemental abundances in the ancient and most metal deficient stars are extremely important because they serve as tests of variable nucleosynthesis pathways and can provide critical inferences of the type of stars that lived and died before them.  The presence of  $r$-process  elements in a handful of  carbon-enhanced metal-poor (CEMP-$r$) stars, which are assumed to  be closely connected to the chemical yield from the first stars, is hard to reconcile with standard neutron star mergers.  Here we show that the production rate of dynamically assembled compact binaries in high-$z$ nuclear star clusters can attain a sufficient high value  to be  a potential viable source of heavy $r$-process material in CEMP-$r$ stars. The predicted frequency of such events in the early Galaxy,  much lower than the frequency of Type II supernovae but with significantly higher  mass  ejected per event, can naturally lead to a high  level of scatter of Eu as observed in CEMP-$r$  stars. 
\end{abstract}


\section{Introduction}

The oldest persisting  stars in the galactic halo serve as laboratories for studies of neutron-capture element synthesis in the early Universe  \citep{Sne08}. Their chemical compositions provide evidence about the identities of the first generations  of stars that lived and died before them \citep{truran2002,Mat14}. $r$-process elements are commonly observed in stars with metallicities $\mathrm{[Fe/H]}\lesssim-3$, indicating that their progenitors must have been relatively swiftly evolving. Of particular significance is the rich $r$-process element composition found in some carbon-enhanced metal-poor (CEMP-$r$) stars \citep[e.g. CS\,22892$-$052;][]{Sne03,Mass2010}. The presence of carbon in these stars  has been closely connected to the chemical yield from the first stars\footnote{Although it is still debated whether these enhancements are related to the star's birth composition, or if its atmosphere  was subsequently polluted by a binary companion \citep[e.g.][]{Sta2014}.} \citep{UmeNom03,CooMad14} whilst the $r$-process must either come from the same  first stars or from a source that acts on a timescale shorter than the time required to form a second generation of stars. 

The neutrino driven wind in Type II supernovae (SNe) has long
been considered a likely site for $r$-process synthesis, based on
both the physical conditions found in early simulations
\citep{woosley1994}, and chemical evolution
considerations in the early Universe
\citep[][]{Arg04,cowan2004}. Later work has shown
that it is difficult to get conditions in the wind which 
produce the $r$-process \citep[e.g.][]{takahashi1994,Qian96}.
Neutron star  mergers  offer a robust  alternative to Type II SNe
\citep{Lattimer74,Fre1999}. The $r$-process nuclei are
robustly synthesized in the matter ejected in such
mergers  \citep{Metzger10,Roberts11,Bauswein13,Grossman13} and
their predicted enrichment history  is  in agreement with
abundance patterns observed in galactic halo stars
\citep{SCR14,Voort 2014}. However,  because evolved  neutron star
binaries  are expected  to merge hundreds to thousands of
millions of years  after birth
\citep{fryer1999,Kal2001,Bel06,Beh2014}, 
a shorter-lived  merging channel  might be required  in order to
explain $r$-process enrichment in CEMP-$r$ stars.

In this {\it Letter}, we explore one such channel by investigating
whether or not sufficient $r$-process material can be synthesized
in the first star clusters formed within the first
$500~\mathrm{Myr}$ after Big Bang via dynamically assembled
merging compact binaries
\citep[][]{lee1993,lee2010,Sam2014}.  To
answer this question, we first estimate the formation rate of
nuclear star cluster in the early Universe as well as their
structural properties, which we use to compute detailed estimates
of the rate of compact object encounters within such systems
(Section \ref{sec:ass}). From the stellar binary merger rate, we then derive the
amount of $r$-process material assembled in the early Universe using
the typical mass production rate per event, which is addressed in
detail for compact binary encounters in Section \ref{sec:rp}.
Discussion and conclusions are presented in Section
\ref{sec:dis}.

\section{Compact Binary Assembly in  The Early Universe}\label{sec:ass}

\subsection{The Formation Rate and Characteristics  of High-\lowercase{$z$} Star Clusters}\label{sec:form}

The basic assumption of our modeling is that compact star clusters can
be formed at the center of high-$z$ dark-matter halos with virial
temperature $T_{\rm vir}\sim(1-2)\times10^4~\mathrm{K}$ and non-zero
but very low metallicity $\log_{10}{(Z/Z_{\sun})}\sim-4$. In fact,
under these conditions disk fragmentation can be suppressed and gas
efficiently funneled by the Toomre instability in the innermost few
parsecs where it forms a star cluster \citep[see][for
details]{devecchi09}.

To estimate  the formation rate of early forming compact and
high-density star clusters, we derive the formation rate of dark matter (DM) halos
meeting the criteria above from the simulations of
\citet{trenti_shull10}, which focus on early chemical enrichment of
three Milky Way-like halos with DM mass $M_{\rm DM}(z=0)\approx10^{12}~\mathrm{M_{\odot}}$. The simulations have been carried out
using $N_{\rm DM}=1024^3$ DM particles in a comoving volume
of $10^{3}~\mathrm{Mpc^3}$, with DM mass resolution of
$3.4\times10^4~\mathrm{M_{\odot}}$. The DM  only simulation
is post-processed with the star formation and chemical enrichment
model described in \citet{trenti_shull10} and \citet{trenti09}, which
includes treatment of metal outflows and radiative feedback in the
Lyman-Werner bands to regulate Population III formation \citep{ts09}.
\begin{figure}[b]
\plotone{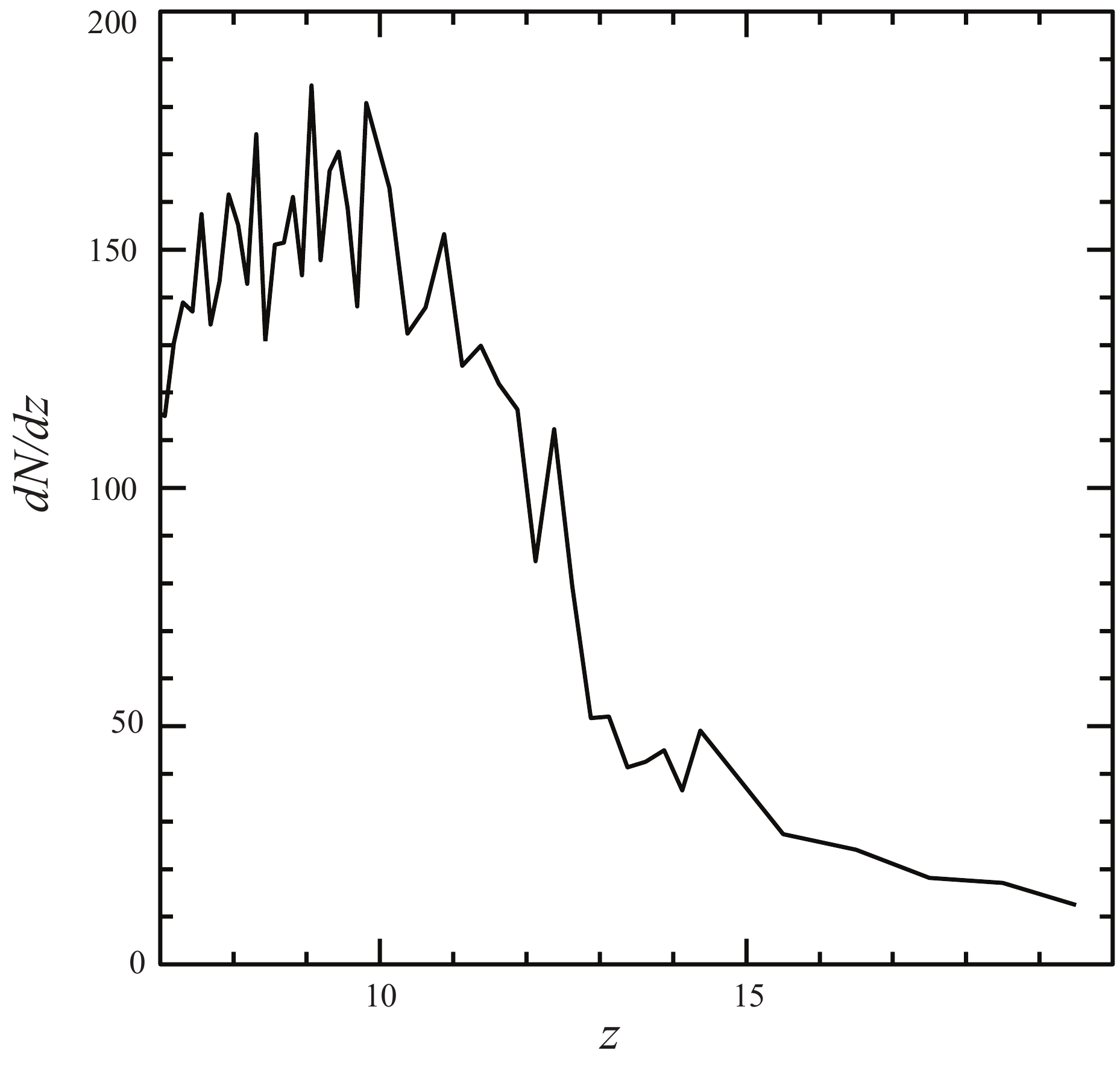}
\caption{
The formation rate for central star clusters as inferred from three Milky Way
like-halos from \citet{trenti_shull10}. Shown are the number of clusters per unit
redshift that will be part of a Milky-Way-like halo at $z=0$,  which are assumed to form  at the
center of dark-matter halos with $T_{\rm vir}\gtrsim10^4~\mathrm{K}$ and $Z\sim10^{-4}Z_{\odot}$.}
\label{fig1}
\end{figure}

Because, as shown in Section~\ref{sec:rates}, the encounter rate in
the core of the first star clusters declines significantly at redshifts $z<10$
(mainly due to a decrease in the central density), we focus only on
systems formed before then, which are shown in Fig.~\ref{fig1}. By
$z=10$, a total of $\langle N_{\rm c}\rangle=515$ clusters are
expected to be formed in a region that will collapse by $z=0$ in a
Milky Way--like halo. This number is higher by a factor of about two
when compared to the formation rate of similar objects in a random
region of the Universe, owing to the enhanced DM  halo
formation rate induced by the over dense environment \citep[see
e.g.][]{trenti_shull10}.

Overall our simplified approach provides a robust estimate of the
formation rate of halos capable of hosting dense star clusters. For
example, feedback at $z>10$ is unlikely to affect gas in halos with
$T_{\rm vir}>10^4$ K. This leaves the leading source of uncertainty in
the details of star formation in such halos, which are very
challenging to model from first principles
\citep[e.g.][]{wise2012}. Thus, as a first characterization, we
describe the internal properties of first stellar clusters at
$z\approx10$ following \citet{devecchi09}. From Fig.~4 in
\citet{devecchi09} we derive the characteristic stellar mass
$M_{\mathrm{ tot}}$ and half-mass radius $r_{\mathrm{hm}}$ for the
central star clusters as a function of their formation redshift. While
the core density during initial stages of cluster evolution is fairly
uncertain and dependent upon the initial conditions, as a basic
estimate, based on direct N-body modeling experiments
\citep{trenti10}, we assume a typical core to half mass radius ratio
$r_{\mathrm{c}}/r_{\mathrm{hm}}=0.1$ and a core to characteristic
stellar mass ratio {$M_{\mathrm{c}}/M_{\mathrm{tot}}=0.04$,
  corresponding to a pre-core collapse cluster with a
  $W_0=8.5$~\citet{king66} profile. With the model (in virial
  equilibrium), we derive the central velocity dispersion from the
  cluster mass and radius:
\begin{equation}
\sigma_{\mathrm{c}}=18.81~\mathrm{km~s^{-1}}\left(\frac{M_{\mathrm{tot}}}{10^5~\mathrm{M_{\odot}}}\right)\left(\frac{1~\mathrm{pc}}{r_{\mathrm{hm}}}\right). 
\end{equation}}

\begin{figure}[t]
\plotone{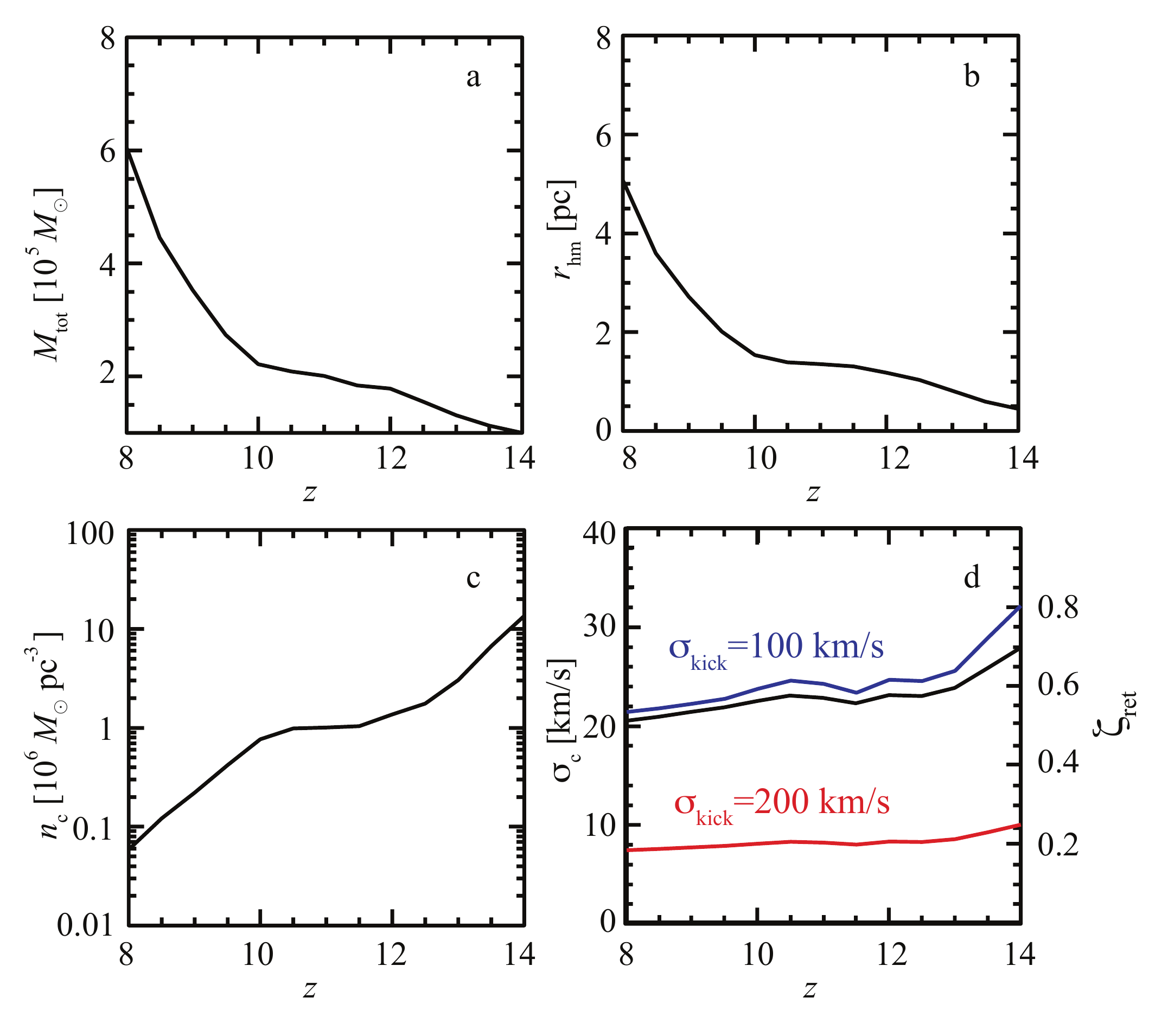}
\caption{The structural properties of high-$z$ stellar clusters as a
  function of the redshift of formation, as predicted by the
  \citet{devecchi09} model for mass and radius, and combined with the
  typical structural parameters from \citet{trenti10} to infer core
  properties. Plotted are the characteristic stellar mass $M_{\rm
    tot}$ (panel a), half-mass radius $r_{\rm hm}$ (panel b), core
  stellar density $n_{\rm c}$ (panel c) and core velocity dispersion
  $\sigma_{\rm c}$ {(panel d; black line). Panel d also shows the
    core retention fraction $\zeta_{\rm ret}$ for natal kicks with
    Maxwellian distribution and dispersion
    $\sigma_{\rm kick}=100~\mathrm{km/s}$ (blue line) and at
    $\sigma_{\rm kick}=200~\mathrm{km/s}$ (red line).}}
\label{fig2}
\end{figure}

Having an estimate for the mass and velocity dispersion of a cluster
(Fig.~\ref{fig2}), the number of neutron stars (NSs) can be calculated
by assuming the star clusters have a {\citet{kroupa01} IMF
  $\xi(m)=\xi_0(m/0.5M_{\odot})^{\alpha(m)}$, with
  $m\in[m_{\mathrm{min}}:m_{\mathrm{max}}]$ and either a cut-off at
  $m_{\mathrm{min}}=0.2M_{\odot}$ or at $m_{\mathrm{min}}=1~M_{\odot}$
  (top-heavy IMF). $m_{\mathrm{max}}=100M_{\odot}$,
  $\alpha(m)=-2.3$ for $m>0.5M_{\odot}$ and $\alpha(m)=-1.3$ for
  $0.2M_{\odot}\leq m\leq0.5M_{\odot}$.} By assuming, for example,
that NSs are produced in the mass range $[8:25]~\mathrm{M_{\odot}}$
\citep{hurley2000}, the fraction of NSs is {$f_{\rm ns}=9\times
  10^{-3}$ for the standard IMF} and $f_{\mathrm{ns}}=5\times10^{-2}$
for the top heavy IMF.  

{Next, we estimate the fraction of these NSs that are retained in the
  cluster core.  The core retention fraction is affected by NS natal
  kicks, binary and mass segregation.  We assume that the progenitors
  of NSs are generally in binaries, since massive stars are expected
  to form with high binarity \citep{krumholz09}, and use the model of
  \citet[][Fig.~12]{pfahl02} to estimate the retention of NSs formed
  in clusters with different escape velocities. We consider a
  Maxwellian kick distribution,
  $p(v)=\sqrt{2/\pi}v^2/\sigma^3_{\rm kick} e^{-v^2/(2\sigma^2_{\rm
      kick})}$,
  with $\sigma_{\rm kick}=100~\mathrm{km/s}$ and
  $\sigma_{\rm kick}=200~\mathrm{km/s}$, which are the fastest kick
  distribution studied by \citet{pfahl02}. Since that study concludes
  that even with their optimistic assumptions they might underestimate
  actual retention fraction (Section 7 in \citealt{pfahl02}), our
  approach is fairly conservative. We assume a cluster escape speed
  $v_{\rm esc}(0)=4.12\sigma_{\mathrm{c}}$ as derived from the King
  density profile.  In what follows we will assume that 50\% of the
  retained NSs are unbound from their parent binary while 50\% remain
  bound, as roughly found by \citet{pfahl02} for the range of
  cluster's escape velocities derived here.  Mass segregation can
  enhance the number of NSs and NS-progenitors in the core relative to
  the cluster mean.  From N-body simulations with stellar evolution
  (MacLeod et al. 2015, in preparation), we see that mass segregation
  of the NS progenitors is efficient (core density increased by
  $\approx4$), but the core density of NSs is increased only by
  $\approx50\%$, owing to the redistribution following natal kick. We
  thus apply this latter correction factor (that is $1.5\times$) to
  map global to core retention fraction ($\zeta_{\rm ret}$), which is
  then shown in panel d of Fig.~\ref{fig2}. $\zeta_{\rm ret}\sim0.2$
  for fast kicks ($\sigma_{\rm kick}=200~\mathrm{km/s}$) at all
  redshifts, while for the distribution with slower kicks
  $\zeta_{\rm ret}\sim0.8$ at $z=14$, decreasing to
  $\zeta_{\rm ret}\sim0.55$ by $z\sim10$. }
  
\subsection{Neutron Star Binary Assembly Rates}\label{sec:rates}
To calculate the encounter rate within the cluster we assume NSs are distributed homogeneously within the core, with fractional
number $f_{\rm ns}$, {total number density $n_{\rm c}$, and
  binary fraction $b_{\rm ns}$}. We further assume that the stars
follow a Maxwellian velocity distribution function with dispersion
$\sigma_{\rm c}$. 
We calculate a fiducial encounter rate $\xi_{\rm ns}$ for individual clusters provided a distance of closest approach $R_{\rm  min}$  \citep[e.g.][]{lee2010}, 

\begin{multline}
  \label{eq:colrate}
 \xi_{\rm ns}= 
  3\times 10^{-3}\;\mathrm{Gyr}^{-1} f_{\rm ns}^2\left(\frac{n_{\rm c}}{10^6\;\mathrm{pc}^{-3}}\right)^2 \left(\frac{r_{\rm c}}{0.1\;\mathrm{pc}}\right)^3 \times \\ \left(\frac{\sigma_{\rm c}}{20\;\mathrm{km\,s}^{-1}}\right)^{-1}  \left(\frac{M_{\rm ns}}{1.4\;\mathrm{M}_\odot}\right)\left(\frac{R_{\rm min}}{10\;\mathrm{km}}\right).
\end{multline}
To estimate the rate of single-single NS encounters, we multiply by the single NS fraction squared, $1/2(1-b_{\rm ns})^2$, where the factor of $1/2$ avoids double counting pairwise encounters. To estimate binary-single encounters, we must include a factor $b_{\rm ns}(1-b_{\rm ns})$.
To form a binary by transferring orbital energy to internal stellar oscillations, we adopt the cross section for tidal capture formalism described in \citet{lee86} and \citet{kimlee99}, which in the case of  NSs give 
\begin{equation}
R_{\rm min}^{\rm tidal}=13.37R_{\rm ns},
\end{equation}
 \citep{lee2010}. To compute the cross section for  binary formation by GW radiation, we follow \citet{lee1993}. That is:
\begin{equation}
R_{\rm min}^{\rm gw}=1458\left(\frac{\sigma_c}{20~\mathrm{km~s^{-1}}}\right)^{-4/7} \mathrm{km}.
\end{equation}
This is significantly larger than  $R_{\rm min}^{\rm tidal}=160(R_{\rm ns}/12{\rm km})\;{\rm km}$. Although, as we show in Section~\ref{sec:rp},  tidal capture encounters eject significantly more $r$-process material. 

\begin{figure}
\plotone{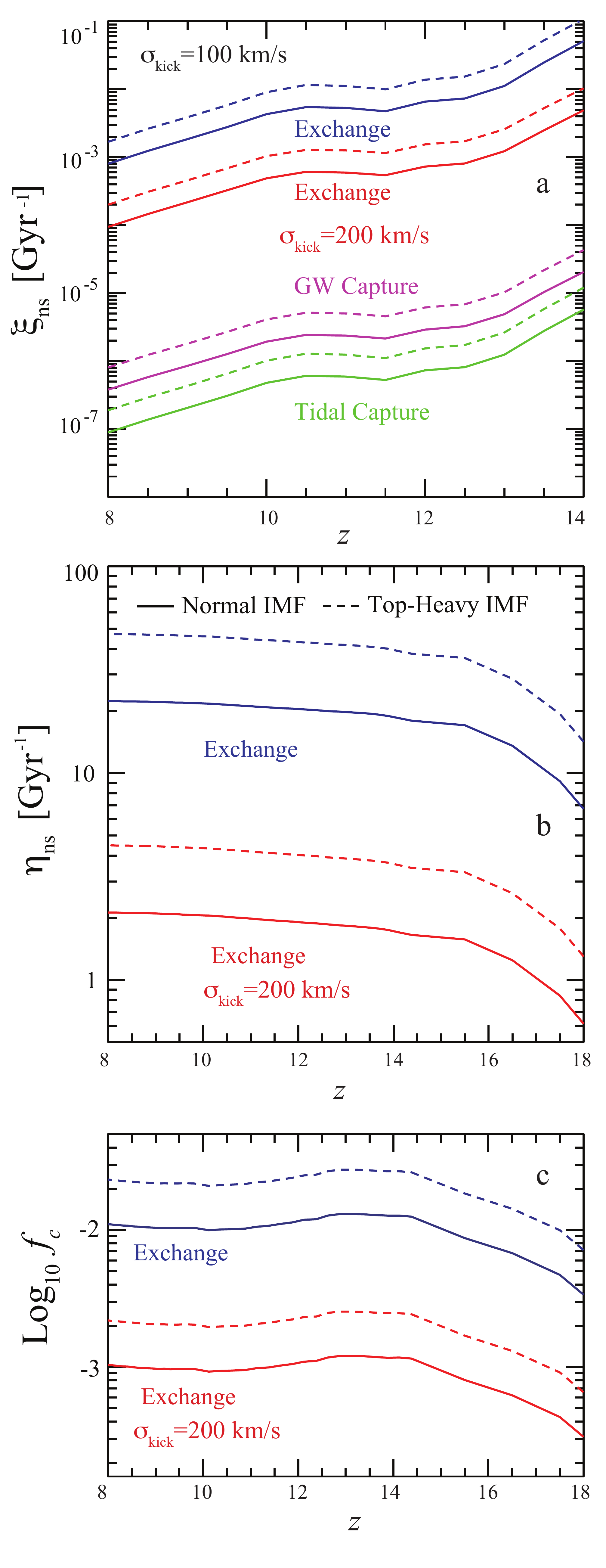}
\caption{The formation rate of compact binaries in the early
  Universe. Panel a: The binary assembly rate $\xi_{\rm ns}$ for
  individual clusters as a function of their formation redshift. The
  rate is higher for the clusters formed at higher redshift, because
  they have higher concentrations.  Panel b: The integrated event
  rates $\eta_{\rm ns}$ for {dynamically formed binaries} as a
  function of $z$. For this we have used the cluster formation history
  plotted in Fig.~\ref{fig1} and then assumed that each cluster is
  active in producing collisions for 0.5Gyr at the rate set by its
  formation redshift. In addition, at $z\geqslant14$ we have assumed a
  constant encounter rate as the \citet{devecchi09} model does not
  extend to higher $z$. This is a conservative assumption as the
  encounter rate increases with redshift.  Panel c: The fraction of
  star clusters $f_{\rm c}$ that experienced a merger as a function of
  redshift.}
\label{fig3}
\end{figure}

Finally, we consider encounters between one binary containing a
  NS with a single NS leading to an exchange, and subsequent formation
  of a NS-NS pair. 
  Using exchange cross sections from \citet{1996ApJ...467..359H}, the cross section for a $1.4M_\odot$ NS to exchange into a binary given an equal mass companion is $\sigma_{\rm ex} = 32.2 {\rm AU}^2 (a/{\rm AU}) (v/20{\rm km~s}^{-1})^{-2}$. 
    If the eccentricity is sufficiently high, the pair
  will merge on a short timescale \citep{Sam2014}.
  To calculate the cross section for the formation of NS binaries with lifetime less than 1 Gyr, 
  we use the lifetime estimate  $t_{\rm life} \approx 2.9 \times 10^{17} (1-e^2)^{7/2}\text{ yr}$ and assume thermal  distribution of eccentricity, $p(e)=2e$, following the exchange \citep{Sam2014}. Thus, for any binary semimajor axis, $a$, there is a critical eccentricity $e_{\rm crit}$ above which newly formed binaries will merge in less than 1 Gyr. 
  The fraction of binaries with $e>e_{\rm crit}$ is thus $1-e_{\rm crit}^2$ and scales as $a^{-8/7}$. The resultant cross section scales weakly with binary semimajor axis, $a^{-1/7}$ \citep[e.g.][~Fig. 15]{Sam2014}.  
  For a representative initial binary population with 
  binary semimajor axis distribution with $p(a) \propto a^{-1}$ and
  $a\in[10~R_\odot:1~\mathrm{AU}]$, the integrated cross-section can
  be expressed as 
  \begin{equation}\label{rmin3body}
  R_{\rm min}^{\rm 3-body}=1.6\times10^6\text{ km},
  \end{equation}
in equation~\ref{eq:colrate}.\footnote{If, instead of equal mass, we assume the binary companion mass is 3 times the NS mass ($4.2 M_\odot$, or a turnoff age of $\approx300$Myr) then the effective exchange cross section is reduced \citep{1996ApJ...467..359H}, but if an exchange occurs, the resultant binary is typically tighter by a factor of $a_{\rm f}\approx m_{\rm cap}/m_{\rm ej}a_{0}$. Both of these effects play a role in determining the cross section for rapidly merging NS binaries. Under these assumptions $R_{\rm min}^{\rm 3-body}=3.0 \times10^5\text{ km}$, a factor of $\approx 5$ smaller than equation \ref{rmin3body}.}

Equipped with {$R_{\rm min}^{\rm tidal}$, $R_{\rm min}^{\rm gw}$,
  and $R_{\rm min}^{\rm 3-body}$,} we can calculate the binary
assembly rate $\xi_{\rm ns}$ for individual clusters
(equation~\ref{eq:colrate}) as a function of their formation redshift,
which, as shown in Fig.~\ref{fig2}, sets their structural
properties.  The rates for {tidal, GW captured and dynamically
  formed binaries} within
individual clusters are given in panel a of Fig.~\ref{fig3}, {where we assume $b_{\rm ns}=0.5$}. As
expected, tidal encounters are {very rare and the rate is
  dominated by binary-single interactions, which have been proposed as
viable merger channel by several studies \citep{Sig1993,Gri2006,Sam2014}}. {In this work we ignore binary-binary encounters and their exchange products, which will represent, at most, a contribution of order unity to the binary-single encounter rate.}

An estimate of the integrated encounter rate $\eta_{\rm ns}$ over all
the Milky Way central star clusters present at high $z$ can then be
made under the assumption that each cluster contributes to the
encounter rate from its redshift of formation until its
dissolution. Conservatively, we assume that dissolution happens
$0.5~\mathrm{Gyr}$ after cluster formation, although some clusters may
well survive for longer and therefore provide additional contribution
to the encounter rate\footnote{Essentially, with this choice we are
  providing the encounter rate at $z>6$ ($t<1~\mathrm{Gyr}$) {and we
    note that compact binaries will survive cluster tidal dissolution
    if that happens before the merger.}}. Our estimate of
$\eta_{\rm ns}$ for {dynamically formed} binaries as a function of $z$
is shown in panel b of Fig.~\ref{fig3}. From this panel it is possible
to compute the total number of mergers (which will follow {within 1
  Gyr} after binary capture) in the Milky Way by integrating the rate
over time.  In panel c of Fig.~\ref{fig3} we also plot for clarity the
fraction of high-$z$ star clusters that experienced a merger as a
function of $z$ derived by dividing the number of expected mergers by
the number of clusters present.  To calculate the total amount of
$r$-process material synthesized in these clusters we need to combine
the rate calculations presented here with an estimate of the mass
production rate per event. It is to this issue that we now turn our
attention.
 
\section{\lowercase{$r$}-Process Synthesis in Merging Binaries}\label{sec:rp}

The  physical conditions that characterized the decompressed
ejecta from compact binary mergers \citep{Lattimer74} are
compatible with the assembly of an $r$-process pattern that is
generally consistent with solar system abundances
\citep{Fre1999}. The most recent numerical studies of
circular (i.e., zero-eccentricity)  compact binary mergers
\citep[e.g.][]{Bauswein13,Hotokezaka2013} shows  that  they eject
about $10^{-3}-10^{-2}~\mathrm{M_\odot}$ of $r$-process material
\citep{lrr07,faber12}.

In contrast to these circular  mergers, eccentric  mergers (i.e., with finite eccentricity)
can  result  in tidal tails that will synthesize significantly
larger masses of $r$-process rich material
\citep{lee2010,rosswog2013}. But in order for them to offer a
consequential  enhancement  to standard circular mergers,
the  synthesized $r$-process mass should be higher by a factor of
about
$R_{\mathrm{min}}^{\mathrm{gw}}/R_{\mathrm{min}}^{\mathrm{tidal}}$
when compared to  circular $r$-process  production. This
requirement can be written as 
\begin{equation}
\label{eq:massrate}
{M_{\rm r-p}^{\mathrm{tidal}}\over M_{\rm
r-p}^{\mathrm{gw}}}\lesssim {R_{\mathrm{min}}^{\mathrm{gw}}\over
R_{\mathrm{min}}^{\mathrm{tidal}}
}=115.9 \left(\frac{\sigma_{\rm c}}{20~\mathrm{km~s^{-1}}}\right)^{-4/7}.
\end{equation}

\begin{figure}[b] \plotone{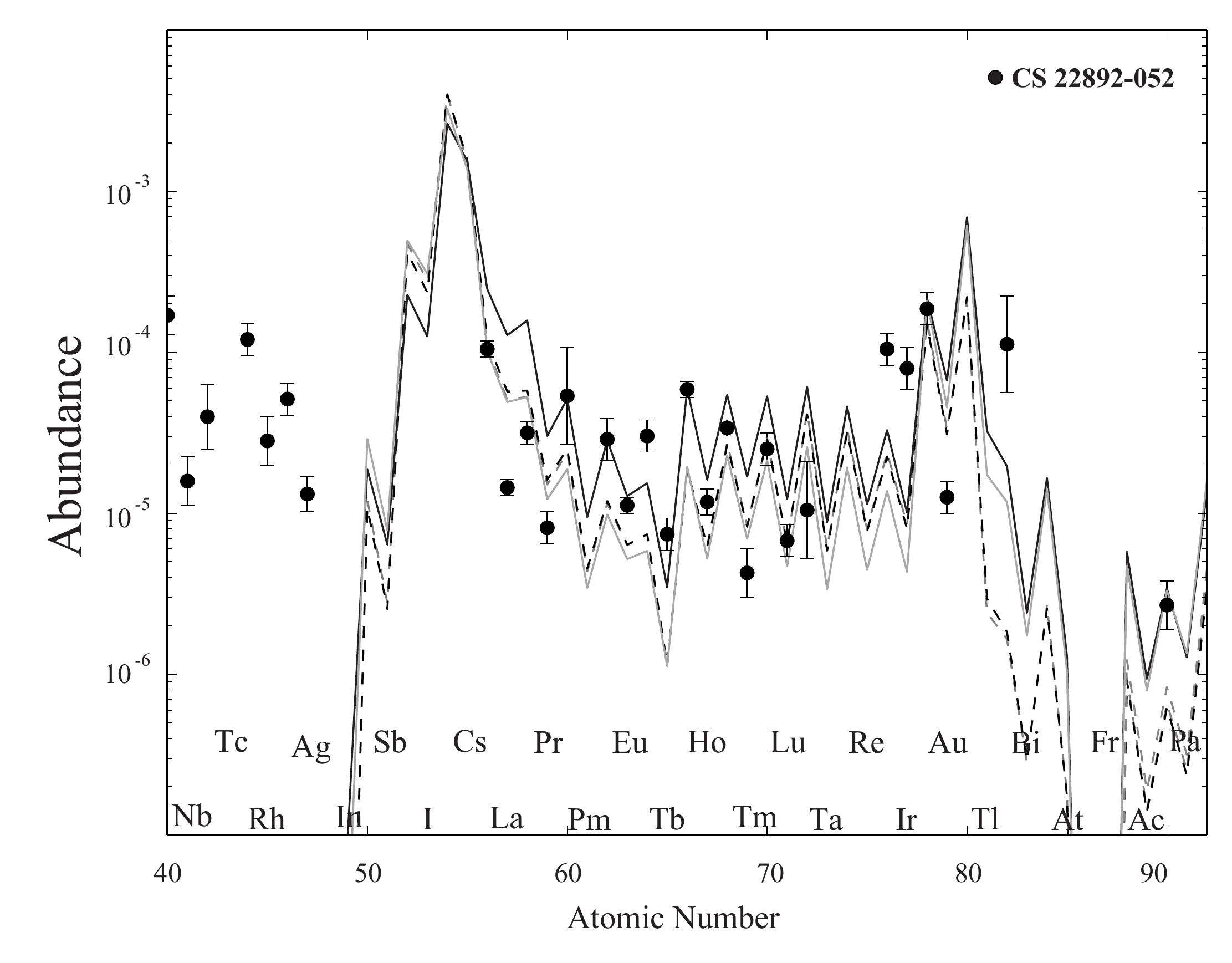} 
\caption{
The abundance patterns calculated for a tidal encounter (solid lines) and a binary merger (dashed lines) compared with the most recent abundance data for the CEMP-$r$ star CS 22892-052 for which [Fe/H]=-3.1.  The merging neutron star binary has a mass ratio $q=0.88$ while the eccentric binary is assembled by the tidal capture of a neutron star by a black hole with  $R_{\mathrm{min}}=2.3R_{\mathrm{ns}}$. The dashed lines show the abundances of two  particle trajectories  in the merging binary as calculated in \citet{Roberts11} while the solid lines correspond to the abundances of two  particle trajectories  in the tidal capture encounter, which have been selected  to define the extent of the abundance variation in the calculations.  The abundance is defined as the number of a particular element per baryon and the CS 22892-052 data has been rescaled for the best fit to the network calculations.} 
\label{fig4} 
\end{figure}

For the two binary members to come together and eventually merge,
they must lose  orbital angular momentum and energy. This can be
achieved through gravitational wave emission, or mass ejection,
or a combination of both.  The tidal interaction provides an efficient mechanism for merging by ejecting a small amount of mass through the formation of tidal tails that carry a great deal of angular momentum to a large radius. In the case of a tidal capture, it is easier to
dynamically unbind matter to infinity and, as a result, the
amount of mass ejected is  larger  than in  a  merger by about
one order of magnitude.  In the case displayed in
Fig.~\ref{fig4}, $M_{\rm r-p}^{\mathrm{tidal}}/M_{\rm
r-p}^{\mathrm{gw}}\approx4$ with $M_{\rm
r-p}^{\mathrm{gw}}=0.19M_\odot$. {Since $M_{\rm
r-p}^{\mathrm{tidal}}/M_{\rm r-p}^{\mathrm{gw}}\ll
R_{\mathrm{min}}^{\mathrm{gw}}/R_{\mathrm{min}}^{\mathrm{tidal}}<R_{\rm min}^{\rm 3-body}/R_{\mathrm{min}}^{\mathrm{tidal}}$,
circular  binary mergers arising from binary exchange\footnote{In the binary channel, only $\approx$1/108 of the exchanges will result in eccentric merging binaries \citep{Sam2014}.}   and GW capture should dominate the $r$-process
mass production in high-$z$ stellar clusters.}  In what follows we
thus neglect the contribution of eccentric  mergers.

In Fig.~\ref{fig4}, the final elemental abundances in
the tidal tails of a merging neutron star binary from \citet{Roberts11}  
are compared with those of
an eccentric binary resulting from the tidal capture of a
neutron star by a black hole, both using the FPS equation of state \citep{shibata05}.  The model results are also
compared with observed elemental abundances of the CEMP-$r$ star CS
22892-052 \citep{Sne03}.  The evolution of the nuclear
composition in the tails is followed using the nuclear reaction
network described in \citet{Roberts11}, but symmetric fission
has been assumed.  All models assumed $Y_e=0.1$.  

The models lack light $r$-process elements due to the
  extremely neutron-rich conditions encountered in the tidal
  ejecta and because we do not follow the disk formed after merger 
  \citep{Just2014}.  For the second and third $r$-process peaks, our 
  abundance distribution is reasonably consistent with that seen in CEMP-$r$
  stars \citep{Sne08}. The origin of the remaining discrepancy is likely
due to the unsettled nuclear physics employed, in particular to the
uncertainties in the fission rates and daughter distributions, and
because $Y_e$ is unconstrained by the merger models used in this
work.

\section{Discussion}\label{sec:dis}

\begin{figure}
\plotone{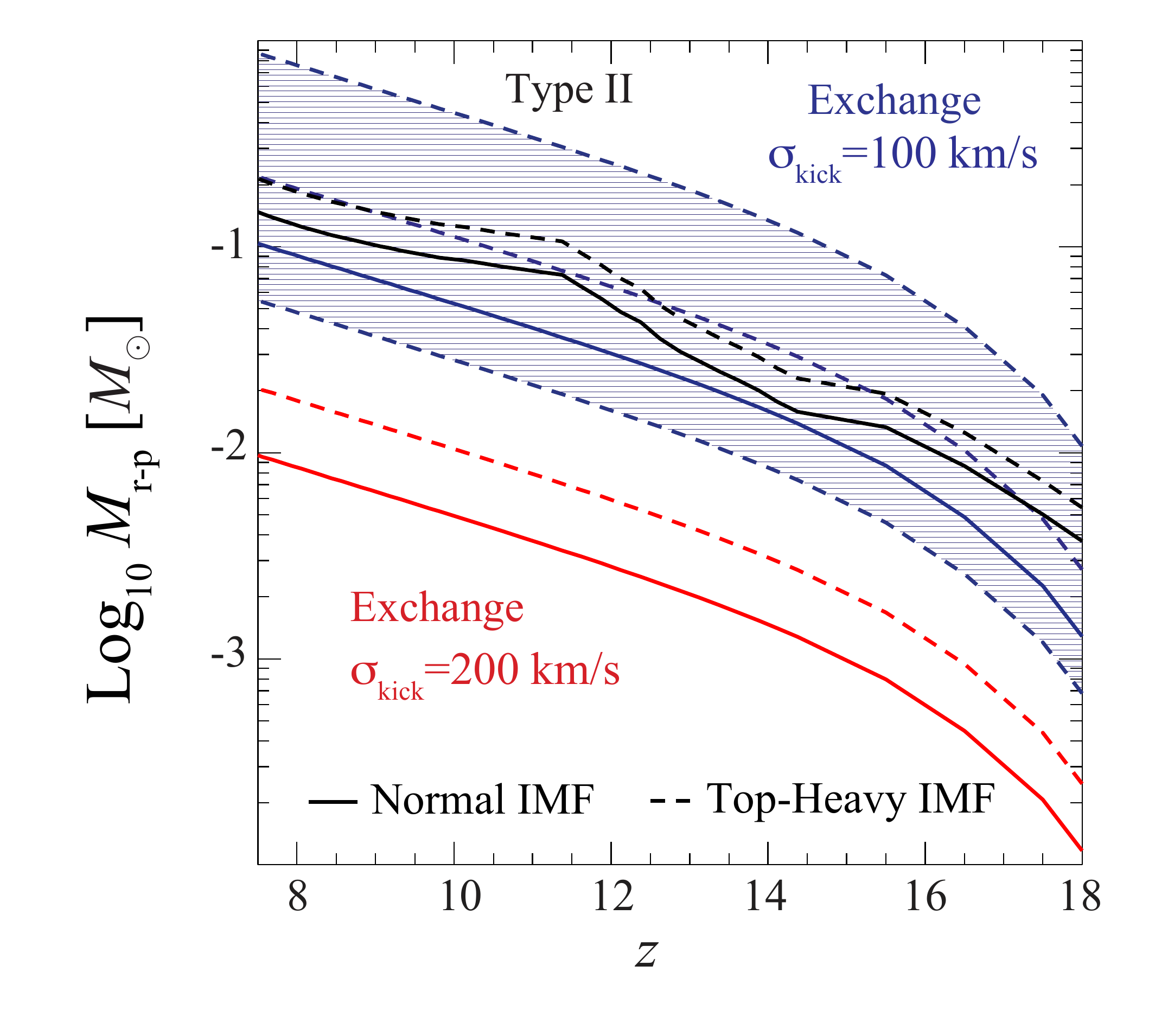}
\caption{Predicted production of $r$-process elements from {compact} binaries in high-$z$ clusters. The red
lines show production by Type II, while the blue lines show production
by {binary} mergers. We assume that the mass of $r$-process material produced per events  is $M_{\mathrm{r-p}}^{\mathrm{sn}}=10^{-5}M_\odot$ and  $5\times10^{-3}M_\odot\lesssim M_{\mathrm{r-p}}^{\mathrm{gw}} \lesssim2\times10^{-2}M_\odot$ (blue shaded region), for Type II and compact  binary mergers respectively. Also,  we assume that only 5\% of all  Type II  events produce $r$-process  consistent with the view that only a small fraction  of the massive stars in the early Universe that produce $\alpha$-elements also yield $r$-process elements \citep{fields2002}.}
\label{fig5}
\end{figure}

Having established the amount of mass of $r$-process material produced
by both mergers and tidal captures we can now proceed to determine the
relevance of dynamically assembled binaries in enriching CEMP-$r$
stars with Eu. Fig.~\ref{fig5} shows the cumulative mass of $r$-process
material as a function of $z$ synthesized in {binary} mergers, as
calculated from the results shown in Fig.~\ref{fig3}.  Typical model
uncertainties of $5\times10^{-3}\mathrm{M_\odot}\lesssim
M_{\mathrm{r-p}}^{\mathrm{gw}} \lesssim2\times10^{-2}\mathrm{M_\odot}$
are shown as the blue shaded region. For comparison we have plotted
the contribution of Type II SN, calculated assuming a mass production
rate of $M_{\mathrm{r-p}}^{\mathrm{sn}}=10^{-5} M_\odot$, which is the
average mass required per event in order for core collapse to be
solely responsible for the $r$-process in the Milky Way
\citep{cowan2004}. In accordance with the higher level of scatter of
[Eu/Fe] in relation to [$\alpha$/Fe], which is most pronounced at
values below [Fe/H]$\approx$-2.0, we have assumed that only a 5\% of
all Type II that produce $\alpha$-elements also yield $r$-process
elements \citep{fields2002}.

Fig,~\ref{fig5} shows that in high-$z$ clusters, albeit with large
uncertainties, dynamically assembled mergers could be {potentially} as
important as Type II SNe. Because they happen only in a few clusters
(Fig.~\ref{fig3}), we expect  high concentrations of $r$-processes in locations where a
binary merger happens compared to a case where Type-II SNe are
assumed to be the main source of the $r$-process. We estimate
  that NS-NS merger blastwaves do not carry enough
  energy and momentum to escape their host minihalo. However, the short halo assembly time
at high-$z$ implies that there is significant mixing because of halo
mergers, leading to  an enhancement of
$r$-process by a  factor $\gtrsim 10$ (for $f_{\rm c} \sim 10^{-2}$) in a few percent of
  the stars formed at $z\gtrsim6$. This can help explain  the large
star-to-star bulk scatter in the concentrations of heavy elements with
respect to the lighter metals in CEMP-$r$ stars \citep{Sne08}.

Studies of element abundances in the oldest and most Fe deficient
stars in our Galaxy have concluded that, because neutron-star
binary timescales are long ($\gtrsim100$ Myr), only SNe
could have contributed to $r$-process synthesis at the earliest
times \citep[e.g.][]{cowan2004}. However, recent hydrodynamical
simulation studies  have shown that assuming a relatively small
time interval between the formation of the Galaxy and appearance
of stars with [Fe/H]$\leq-3$ is not accurate \citep{SCR14,Voort 2014} and
that in fact  neutron star mergers offer a favorable
alternative to Type II SNe as the $r$-process site. For
CEMP-$r$ stars, which are considered the
chemical descendants of Population III stars
\citep{UmeNom03,CooMad14}, any $r$-process enhancement must
happen on a timescale shorter than that of pollution from a
second generation of stars. Dynamically assembled
binaries resulting in a  short-lived merger can fit these
requirements better than the standard  compact mergers, whose
formation timescale is determined by binary stellar evolution and
is estimated to be $\gtrsim100$ Myrs
\citep{fryer1999,Kal2001,Bel06,Beh2014}. Their production rate
in high-$z$ clusters, as calculated in Section~\ref{sec:rates},
could attain a sufficiently high value  for  compact binary mergers
to  still be  a potentially viable source of heavy $r$-process
material in CEMP-$r$ stars. 

\acknowledgments 
We thank  S. Shen, R. Cooke, E. Kirby, C. Miller, M. Rees and  S. Rosswog  for insightful discussions  as well as the editor and  referees for insightful suggestions.  We acknowledge financial support from the Packard Foundation, NSF (AST0847563), UCMEXUS (CN-12-578) and  the Einstein Fellowship (LR).


\begin{thebibliography}{}

\bibitem[\protect\citeauthoryear{Argast et al.}{2004}]{Arg04}
Argast D., Samland M., Thielemann F.-K., Qian Y.-Z., 2004, A\&A, 416, 997

\bibitem[{{Bauswein} {et~al.}(2013){Bauswein}, {Goriely}, \&
  {Janka}}]{Bauswein13}
{Bauswein}, A., {Goriely}, S., \& {Janka}, H.-T. 2013, \apj, 773, 78

\bibitem[Behroozi et al.(2014)]{Beh2014} Behroozi, P.~S., 
Ramirez-Ruiz, E., \& Fryer, C.~L.\ 2014, arXiv:1401.7986 

\bibitem[\protect\citeauthoryear{Belczynski et al.}{2006}]{Bel06}
Belczynski K., Perna R., Bulik T., Kalogera V., Ivanova N., Lamb D.~Q., 2006, ApJ, 648, 1110

\bibitem[\protect\citeauthoryear{Cooke \& Madau}{2014}]{CooMad14}
Cooke R., Madau P., 2014, ApJ submitted, arXiv:1405.7369

\bibitem[Cowan \& Thielemann(2004)]{cowan2004} Cowan, J.~J., \& Thielemann, F.-K.\ 2004, Physics Today, 57, 47 

\bibitem[Devecchi \& Volonteri(2009)]{devecchi09} Devecchi, B. \&
  Volonteri, M. 2009, \apj, 694, 302


\bibitem[Faber 
\& Rasio(2012)]{faber12} Faber, J.~A., \& Rasio, F.~A.\ 2012, Living Reviews in Relativity, 15, 8 


\bibitem[Fields et al.(2002)]{fields2002} Fields, B.~D., Truran, 
J.~W., \& Cowan, J.~J.\ 2002, \apj, 575, 845 


\bibitem[Freiburghaus et al.(1999)]{Fre1999} Freiburghaus, C., 
Rosswog, S., \& Thielemann, F.-K.\ 1999, \apjl, 525, L121 

\bibitem[Fryer et al.(1999)]{fryer1999} Fryer, C.~L., Woosley, 
S.~E., \& Hartmann, D.~H.\ 1999, \apj, 526, 152 

\bibitem[Grindlay et al.(2006)]{Gri2006} Grindlay, J., 
Portegies Zwart, S., \& McMillan, S.\ 2006, Nature Physics, 2, 116 


\bibitem[{{Grossman} {et~al.}(2014){Grossman}, {Korobkin}, {Rosswog}, \&
  {Piran}}]{Grossman13}
{Grossman}, D., {Korobkin}, O., {Rosswog}, S., \& {Piran}, T. 2014, \mnras,
  439, 757
 
 
\bibitem[Heggie et al.(1996)]{1996ApJ...467..359H} Heggie, D.~C., Hut, P., 
\& McMillan, S.~L.~W.\ 1996, \apj, 467, 359 
  
  \bibitem[Hotokezaka et al.(2013)]{Hotokezaka2013} Hotokezaka, K., 
Kiuchi, K., Kyutoku, K., et al.\ 2013, \prd, 87, 024001 

\bibitem[Just et al.(2014)]{Just2014} Just, O., Bauswein, A., 
Ardevol Pulpillo, R., Goriely, S., \& Janka, H.-T.\ 2014, arXiv:1406.2687   

\bibitem[Kalogera et al.(2001)]{Kal2001} Kalogera, V., Narayan, 
R., Spergel, D.~N., \& Taylor, J.~H.\ 2001, \apj, 556, 340 
  
\bibitem[Kim \& Lee(1999)]{kimlee99}Kim, S.~S. \& Lee, H.~M. 1999,
  A\&A, 347, 123

\bibitem[King(1966)]{king66} King, I.~R. 1966, \aj, 71, 64

\bibitem[Kochanek(1992)]{koch92} Kochanek, C.~S.\ 1992, \apj, 385, 604

\bibitem[Kroupa(2001)]{kroupa01} Kroupa, P. 2001, \mnras, 322, 221

\bibitem[Krumholz et al.(2009)]{krumholz09} {Krumholz}, M.~R. et
  al. 2009, Science, 323, 754

\bibitem[Hurley et al.(2000)]{hurley2000} Hurley, J.~R., Pols, 
O.~R., \& Tout, C.~A.\ 2000, \mnras, 315, 543 

\bibitem[{{Lattimer} \& {Schramm}(1974)}]{Lattimer74}
{Lattimer}, J.~M., \& {Schramm}, D.~N. 1974, \apjl, 192, L145


\bibitem[Lattimer et al.(1977)]{Lat1977} Lattimer, J.~M., 
Mackie, F., Ravenhall, D.~G., \& Schramm, D.~N.\ 1977, \apj, 213, 225 

\bibitem[Lee(1993)]{lee1993} Lee, M.~H.\ 1993, \apj, 418, 147 

\bibitem[Lee \& Ostriker(1986)]{lee86}Lee, H.~M., \& Ostriker,
  J.~P. 1986, \apj, 310, 176

\bibitem[Lee 
\& Ramirez-Ruiz(2007)]{lrr07} Lee, W.~H., \& Ramirez-Ruiz, E.\ 2007, New Journal of Physics, 9, 17 

\bibitem[Lee et al.(2010)]{lee2010} Lee, W.~H., Ramirez-Ruiz, 
E., \& van de Ven, G.\ 2010, \apj, 720, 953 

\bibitem[Masseron et 
al.(2010)]{Mass2010} Masseron, T., Johnson, J.~A., Plez, B., et al.\ 2010, \aap, 509, AA93 

\bibitem[\protect\citeauthoryear{Matteucci et al.}{2014}]{Mat14}
Matteucci F., Romano D., Arcones A., Korobkin O., Rosswog S., 2014, MNRAS, 438, 2177

\bibitem[{{Metzger} {et~al.}(2010){Metzger}, {Mart{\'{\i}}nez-Pinedo},
  {Darbha}, {Quataert}, {Arcones}, {Kasen}, {Thomas}, {Nugent}, {Panov}, \&
  {Zinner}}]{Metzger10}
{Metzger}, B.~D., {et~al.} 2010, \mnras, 406, 2650

\bibitem[McLaughlin \& van del Marel(2005)]{mclaughlin05} McLaughlin
  D.~E. \& van der Marel, R.~P. 2005, \apjs, 161, 304

\bibitem[Pfahl et al.(2002)]{pfahl02} Pfahl E. et al. 2002, \apj, 573, 283

\bibitem[Qian \& Woosley(1996)]{Qian96} Qian, Y.-Z., \& Woosley,
S.~E.\ 1996, \apj, 471, 331 

\bibitem[{{Roberts} {et~al.}(2011){Roberts}, {Kasen}, {Lee}, \&
  {Ramirez-Ruiz}}]{Roberts11}
{Roberts}, L.~F., {Kasen}, D., {Lee}, W.~H., \& {Ramirez-Ruiz}, E. 2011, \apjl,
  736, L21

\bibitem[Rosswog et al.(2013)]{rosswog2013} Rosswog, S., Piran, T., 
\& Nakar, E.\ 2013, \mnras, 430, 2585 

\bibitem[Samsing et al.(2014)]{Sam2014} Samsing, J., MacLeod, 
M., \& Ramirez-Ruiz, E.\ 2014, \apj, 784, 71 

\bibitem[\protect\citeauthoryear{Shen et al.}{2014}]{SCR14}
Shen S.,  Cooke R.~J., Ramirez-Ruiz E., Madau P., Mayer L., Guedes J., 2014, arXiv:1407.3796

\bibitem[Sigurdsson 
\& Phinney(1993)]{Sig1993} Sigurdsson, S., \& Phinney, E.~S.\ 1993, \apj, 415, 631 

\bibitem[Schenk et al.(2008)]{Schenk08} Schenk, O., Bollhofer, 
M., \& Romer, R.~A.\ 2008, SIAM Review, 50, 91 

\bibitem[Shibata et al.(2005)]{shibata05} Shibata, M., Taniguchi, K., \& Uryu, K. 2005, Phys. Rev. D 71, 084021

\bibitem[Sneden et al.(2003)]{Sne03} Sneden, C., Cowan, 
J.~J., Lawler, J.~E., et al.\ 2003, \apj, 591, 936 

\bibitem[\protect\citeauthoryear{Sneden, Cowan, \& Gallino}{2008}]{Sne08}
Sneden C., Cowan J.~J., Gallino R., 2008, ARA\&A, 46, 241

\bibitem[Starkenburg et al.(2014)]{Sta2014} Starkenburg, E., 
Shetrone, M.~D., McConnachie, A.~W.,  \& Venn, K.~A.\ 2014, \mnras, 441, 1217 

\bibitem[Takahashi et al.(1994)]{takahashi1994} Takahashi, K., Witti, J., \& Janka, H.-T.\ 1994, \aap, 286, 857 

\bibitem[Trenti et al.(2010)]{trenti10} Trenti, M., Vesperini, E. \&
  Pasquato, M. 2010, \apj, 708, 1598

\bibitem[Trenti \& Shull(2010)]{trenti_shull10} Trenti, M. \& Shull,
  J.~M. 2010, \apj, 712, 435

\bibitem[Trenti, Stiavelli \& Shull(2009)]{trenti09} Trenti, M.,
  Stiavelli, M. \& Shull, J.~M. 2009, \apj, 700, 1672

\bibitem[Trenti \& Stiavelli(2009)]{ts09} Trenti, M. \& Stiavelli,
  M. 2009, \apj, 694, 879

\bibitem[Truran et al.(2002)]{truran2002} Truran, J.~W., Cowan, 
J.~J., Pilachowski, C.~A., \& Sneden, C.\ 2002, \pasp, 114, 1293 

\bibitem[\protect\citeauthoryear{Umeda \& Nomoto}{2003}]{UmeNom03}
Umeda H., Nomoto K., 2003, Nature, 422, 871

\bibitem[van de Voort et al.(2014)]{Voort 2014} van de Voort, F., 
Quataert, E., Hopkins, P.~F., Keres, D., 
\& Faucher-Giguere, C.-A.\ 2014, arXiv:1407.7039 

\bibitem[Wise et al.(2012)]{wise2012} Wise, J.~H., Turk, M.~J., 
Norman, M.~L., \& Abel, T.\ 2012, \apj, 745, 50 

\bibitem[Woosley et al.(1994)]{woosley1994} Woosley, S.~E., Wilson, 
J.~R., Mathews, G.~J., Hoffman, R.~D., 
\& Meyer, B.~S.\ 1994, \apj, 433, 229 

\end{thebibliography}
\end{document}